\documentclass[12pt,preprint]{aastex}

\slugcomment{Submitted to Ap. J. }

\shorttitle{POLIMERIZATION OF HCN IN CRL618 }
\shortauthors{Pardo et al.}

\begin{document}


\title{OBSERVATIONAL EVIDENCE OF THE FORMATION OF CYANOPOLYYNES IN CRL618 
THROUGH THE POLIMERIZATION OF HCN}



\author{Juan R. Pardo, Jos\'e Cernicharo \& Javier R. Goicoechea$^{1}$}
\affil{IEM - Departamento de Astrof\'{\i}sica Molecular 
              e Infrarroja, CSIC, Serrano 121, E-28006 Madrid, Spain. \\
$^{1}$Present address: LERMA-\'Ecole Normale Sup\'erieure (UMR 8112 du CNRS), 
24 Rue Lhomond, 75231 Paris, France.}
\email{pardo,cerni,javier@damir.iem.csic.es}


\begin{abstract}
The abundance ratio of consecutive members of the cyanopolyynes family has been 
explored in CRL618 using data acquired in a complete line survey covering the frequency 
range 81-356 GHz. The J$_{up}$ range explored for the different molecules is  
the following: 1 to 4 for HCN and HNC, 9 to 39 for HC$_3$N, 31 to 133 for HC$_5$N, and 72 to 85 
for HC$_7$N (not detected beyond J$_{up}$=85). The lowest vibrationally excited state 
of HC$_7$N ($\nu_{15}$ at 62 cm$^{-1}$) has been tentatively detected. Data analysis 
has been performed by extending our previous geometrical and radiative transfer model of the 
slowly expanding envelope (SEE) surrounding 
the compact central continuum source of CRL 618, that was established from the study of rotational 
lines in several vibrationally excited states of HC$_3$N. The new lines analyzed here 
require to model the high velocity wind (HVW) component and the colder circumstellar 
gas, remnant of the AGB phase of CRL618. The derived HC$_3$N/HC$_5$N 
and HC$_5$N/HC$_7$N abundance ratios from this set of uniformly calibrated lines 
are between $\sim$ 3 and 6 in the different regions, similar to standard 
values in the CSM and ISM, and consistent with previous estimates obtained 
from ISO observations and chemical models. 
However, the abundance ratios of HC$_3$N, HC$_5$N 
and HC$_7$N with respect to HCN are at least two orders of magnitude larger than those typical 
for AGB C-rich stars, such as IRC+10216. This fact indicates that, in the short transition 
toward the Planetary Nebula phase, HCN is quickly reprocessed into longer cyanopolyyne 
chains. A similar behavior was previously found in this object for the polyacetylenic chains 
(C$_{2n}$H$_2$).


\end{abstract}




\section{Introduction}

CRL 618 is a C-rich protoplanetary nebula (PPNe). Its chemical richness and 
its complex morphology with a central B0 star surrounded by an ultracompact 
H{\small II} region, and a thick molecular and dusty envelope (Bujarrabal et al. 1988) 
with different outflows (Cernicharo et al. 1989, hereafter CER89), are now well 
known thanks to many detailed observational studies. Detailed 
chemical models for this object have been developed (Cernicharo 2004, hereafter 
CER04) indicating the rapid evolution of the central star and its influence on 
the circumstellar ejected material. Changes seem to occur at short 
-almost human- time scales. The most recent observational results on the morphology are the 
interferometric observations presented in S\'anchez-Contreras \& Sahai, 2004. 
Concerning the chemical composition, a line survey from 80 to 275 GHz has been 
recently completed (Cernicharo et al. 2005).

After the study of the physical conditions in the slowly expanding envelope (SEE) around 
the ultracompact H{\small II} region of CRL 618 using vibrationally excited states of HC$_3$N 
(Pardo et al. 2004, hereafter PAR04), we focus now on the entire cyanopolyynes family: 
HCN, HNC, HC$_3$N, HC$_5$N and HC$_7$N (longer members have not been detected in our survey).


HC$_5$N is the second molecule, after HC$_3$N, in terms of total number of lines detected 
in the millimeter spectrum of CRL 618. It is a linear molecule with a rotational constant 
of 1331.330 MHz and a dipole moment of 4.33 Debyes, first discovered in space by Little 
et al. (1978). The two lowest energy doubly degenerated bending 
modes ($\nu_{11}$ and $\nu_{10}$) have an energy of 105 and 230 cm$^{-1}$ 
respectively (Yamada et al., 2004).  On the other hand, HC$_7$N was first 
discovered in space by Kroto et al. 
(1978). Its rotational constant and dipole moment are 564.0011 MHz and 4.82 Debyes 
respectively. HC$_3$N/HC$_5$N and HC$_5$N/HC$_7$N abundance ratios between 1.4 to 3.0 were 
found in molecular clouds since the earliest studies (Snell et al. 1981). Finally, HCN, 
with a dipole moment and rotational constant of 2.98 Debyes and 44315.976 MHz respectively, 
is one of the most abundant molecules in the interstellar and circumstellar media. In many 
cases HCN lines are optically thick so that observations of isotopologues or 
vibrationally excited states are preferred to probe molecular clouds and circumstellar envelopes. 
{\bf The other two molecules discussed in this paper have the following dipole moments and rotational 
constants: 3.73 Debyes and 4549.059 MHz (HC$_3$N), and 2.98 Debyes and 45331.7845 MHz (HNC).}

The goal of the study conducted in this paper is to get abundance ratios of the 
different members of the cyanopolyynes family in CRL 618 and to compare them with those 
observed in prototypical AGB objects such as IRC+10216. In addition, by 
using new polar species we aim at improving and extending our previous model of CRL618 
(see PAR04). This is another step toward a final model for the whole millimeter wave 
spectrum of this source. The observational procedure and SEE model developments 
were described in detail in PAR04, we only give here a short summary in sections 
\ref{secobs} and \ref{model}. The analysis of the HCN, {\bf HNC}, $v$=0 HC$_3$N, HC$_5$N, and 
HC$_7$N data, with 
the extension of the model, is presented in section \ref{results}. Comparisons 
of the results with chemical model predictions, 
and other discussions, are presented in section \ref{chemicalmodels}. The work 
is summarized in section \ref{summary}.

\begin{table*}
\caption[]{Observational parameters (velocity centroid, width and integrated area) obtained from gaussian 
fits of the absorption and emission components of selected HC$_{5}$N lines {\bf in CRL618}. See 
{\bf beginning of} section  \ref{results} for {\bf more details}.}
\label{tb:lineparameters}
\begin{center}
\begin{tabular}{ccrrrrrrrr}
 & & & \multicolumn{3}{c}{Absorption} & & \multicolumn{3}{c}{Emission} \\
\cline{4-6} \cline{8-10}
 J$_{up}$ & $v$ state & \multicolumn{1}{c}{$\nu$} & \multicolumn{1}{c}{v}  & \multicolumn{1}{c}{$\Delta$v} & \multicolumn{1}{c}{A}  & &
  \multicolumn{1}{c}{v} & \multicolumn{1}{c}{$\Delta$v} & \multicolumn{1}{c}{A} \\
  & $v$  & \multicolumn{1}{c}{(GHz)} & (km$\cdot$s$^{-1}$) & (km$\cdot$s$^{-1}$) & K$\cdot$km$\cdot$s$^{-1}$ & &
  (km$\cdot$s$^{-1}$) & (km$\cdot$s$^{-1}$) & (K$\cdot$km$\cdot$s$^{-1}$) \\
\hline
33 & $v$=0 & 87.86363  & -20.3 & 18.9 & 2.7 & & -43.3 & 23.3 & -0.87 \\ 
33 & $\nu_{11}$ 1$^-$ & 88.00874  & -20.4 & 10.0 & 0.29 & & -33.1 & 8.6 & -0.29 \\
33 & $\nu_{11}$ 1$^+$ & 88.08534  & -22.6 & 23.2 & 0.54 & & -34.7 & 14.1 & -0.55 \\
33 & 2$\nu_{11}$ 0 & 88.22163  & -21.1 &  6.8 & 0.09 & & -30.5 & 16.9 & -0.17 \\
33 & 2$\nu_{11}$ 2$^-$ & 88.23055 & -24.8 & - & - & & -31.2 & - & - \\
33 & 2$\nu_{11}$ 2$^+$ & 88.24034 & -23.2 & 5.1 & 0.09 & & -33.5 & 5.9 & -0.08 \\
33 & $\nu_{10}$ 1$^+$ & 88.04187 &  &  &  & & -34.7 & 6.7 & -0.05 \\
33 & 3$\nu_{11}$ 1$^-$ & 88.33509 & & & & & -31.2 & 4.9 & -0.05 \\
33 & 3$\nu_{11}$ 3$^+$ & 88.41807 & & & & & -28.5 & 7.2 & -0.16 \\
33 & 3$\nu_{11}$ 1$^+$ & 88.48921 & & & & & -31.4 &  7.2 &  -0.12 \\
34 & $v$=0 & 90.52589 & -21.1 & 19.1 & 4.0 & & -38.4 & 34.1 & -1.9 \\
35 & $v$=0 & 93.18813 & -21.1 & 19.3 & 4.2 & & -37.3 & 32.7 & -2.6 \\
35 & $\nu_{11}$ 1$^-$ & 93.34201 & -22.5 & 8.4 & 0.49 & & -33.0 & 18.7 & -0.89 \\
35 & $\nu_{11}$ 1$^+$ & 93.42323 & -23.2 & 22.2 & 0.88 & & -33.5 & 12.9 & -0.69 \\
35 & 2$\nu_{11}$ 0 & 93.56650 & -21.7 & & & & -32.0 & 8.2 & -0.30 \\
\hline
52 & $v$=0 & 138.44167 & {\bf -21.3} &  19.8 & 3.6 & & {\bf -38.0} & 17.7 & -1.3 \\
58 & $v$=0 & 154.41110 & {\bf -21.3} &  14.8 & 2.8 & & {\bf -38.0} & 20.8 & -1.2 \\
59 & $v$=0 & 157.07253 & {\bf -21.3} &  16.1 & 2.9 & & {\bf -38.0} & 19.9 & -1.2 \\
59 & $\nu_{11}$ 1$^-$ & 157.33140 & -23.8 & 13.3 & 1.6 & & -28.3 & 17.6 & -1.2 \\
59 & $\nu_{11}$ 1$^+$ & 157.46775 & -23.7 & 14.5 & 0.92 & & -31.8 & 8.6 & -0.47  \\
59 & 2$\nu_{11}$ 0 & 157.67636 & -19.1 & 16.6 & 0.43 & & -29.9 & 6.8 & -0.19 \\
59 & 2$\nu_{11}$ 2$^-$ & 157.72678 & -21.0 & 14.0 & 0.47 & & -31.2 & 9.7 & -0.28 \\
59 & $\nu_{10}$ 1$^+$ & 157.39079 & -20.0 & 13.2 & 0.32 & & & & \\
\hline
96 & $v$=0 & 255.50940 & -24.6 & 10.1 & 0.94 & & & & \\
96 &  $\nu_{11}$ 1$^-$ & 255.92839 & -23.6 & 6.1 & 0.27 & & & & \\
98 & $v$=0 & 260.82794 & -21.6 & 10.9 & 0.88 & & & & \\
\hline
\end{tabular}
\end{center}
\end{table*}

\section{Observations}
\label{secobs} 
The observations presented in this paper are part of two line surveys of 
CRL618. The first one, now complete, has been carried out with the IRAM-30m telescope 
from 1994 to 2002 (Cernicharo et al. 2005, hereafter CER05) with a frequency 
coverage of 81-279 GHz except for frequencies with high atmospheric opacity 
around 119 and 183 GHz. The second one, still on-going but complete for the species 
we focus on in this work, is performed with the CSO telescope between 280 and 356 GHz. {\bf For 
both instruments the frequency resolution used has been around 1 MHz with $\sim$0.5 GHz wide 
spectrometers.}  
 The J$_{up}$ range covered 
for rotational transitions of HCN {\bf and HNC} is 1 to 4, for HC$_{3}$N 9 to 39, for HC$_{5}$N 31 to 133, 
and for HC$_{7}$N it is 71 to 316, although the lines are not detected beyond J$_{up}$=85. See 
PAR04 for details on the observational procedures at the two facilities.

\section{Previous Model Summary}
\label{model} 
A model has been developed to study the physical conditions that explain the 
whole observed millimeter wave spectrum of CRL618. The precise knowledge of the 
spectral behavior of the continuum emission (see PAR04) is 
paramount since it plays a very important role in the observed line profiles 
(specially in the emission-to-absorption line ratio).  
The model consists of a gas envelope around a central continuum source, considered 
spherical with size and effective temperature adjusted to fit the continuum IRAM-30m data 
(PAR04) and assumed to keep the same spectral index in the range 280-356 GHz (CSO observations). 
The description of the gas component allows different shells with different physical conditions 
to be included in the model. The geometry is not restricted 
to be spherical. Elongated shells with both radial and azimuthal velocity components 
can be defined.

Since the observations reveal that different gas regions are traced by different 
molecular species, we proceed in our analysis in different steps. 
The first one {\bf has been to} study those species arising from the inner $\sim$1.5'' SEE, 
with outflow velocities of $\sim$ 10 to 15 km$\cdot$s$^{-1}$, 
turbulence velocities around 3.5 km$\cdot$s$^{-1}$, and temperatures 
in the range 250-275 K. {\bf The full description of the SEE model is given in 
PAR04. The extension of the SEE model to include both the colder and outer circumstellar 
gas (necessary to explain the rotational lines in the ground vibrational state 
of HC$_3$N, HC$_5$N and HC$_7$N) and the high velocity wind component (seen in lines 
of some abundant species, {\bf including $v$=0 HC$_3$N}) is one of the targets of this work. This 
is discussed in detail in the next section.}

\section{Analysis and Discussion}
\label{results}

The first step of the analysis has been to perform gaussian fits of selected HC$_{5}$N lines in the 3, 2, and 
1.3 mm windows. A summary {\bf for a set of lines representative of the different vibrational 
states and frequency ranges} is presented in Table \ref{tb:lineparameters} where the areas are expressed in 
K$\cdot$km$\cdot$s$^{-1}$ (T$_A^*\cdot$v$_{LSR}$), parameters that could not be determined 
are marked with a dash, and parameters that 
had to be fixed in order to obtain a reasonable fit are highlighted in bold face. Also, when the velocity, width and 
area of one component are left blank in the table it means that the component is not detected in the particular 
transition. These observational results already reveal an 
interesting behavior that is crucial for the analysis: lines in the $v$=0 state have 
absorption and emission components that are in general 15 to 25 km$\cdot$s$^{-1}$ wide at 3 and 2 mm, with 
the absorption centered at around -37 to -42 km$\cdot$s$^{-1}$, whereas lines in vibrationally excited states, 
in the same spectral region, appear narrower ($\sim$ 7 to 15 km$\cdot$s$^{-1}$) with the absorption 
component (sometimes the only one detected) centered at $\sim$ -28 to -33 km$\cdot$s$^{-1}$. These  
differences tend to disappear as frequency increases and the line profiles in the $v$=0 state are dominated 
by the emission component only. These facts suggest that the rotational lines in the ground state of HC$_{5}$N are mostly 
formed in a region with quite different physical conditions than the one responsible for the lines in 
vibrationally excited states. This is carefully explored in this section. 

\begin{figure}[t]
\includegraphics[angle=0,width=12cm]{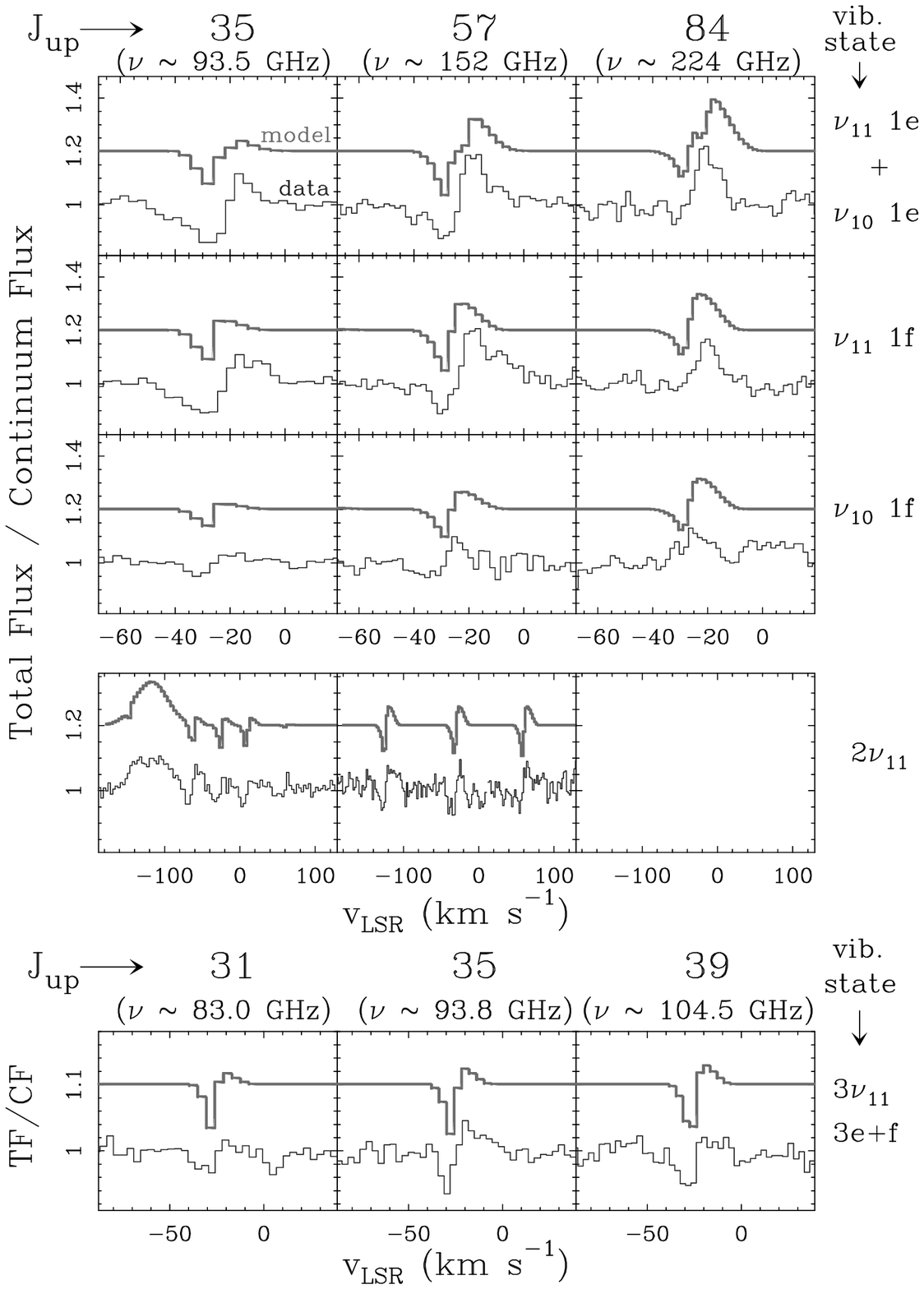}
\caption{Sample of detected lines in vibrationally excited states of HC$_5$N toward 
CRL 618. Model results, {\bf shifted by 0.2$\cdot$F$_c$ (F$_c$: Continuum Flux) for all 
vibrational states except for 3$\nu_{11}$ (0.1$\cdot$F$_c$ shift)}, are shown at 
the {\bf velocity} resolution of the data 
{\bf ($\Delta v=c\cdot\frac{\Delta\nu}{\nu}$, with $\Delta\nu\simeq$1 MHz} for all spectra) 
for a HC$_3$N/HC$_5$N value of 3. 
All other physical conditions are as determined in PAR04 in the SEE (the only region considered). The extra 
feature that appears in the 2$\nu_{11}$ J$_{up}$=35 spectrum is a blending of the $\beta$51 recombination line 
of atomic hydrogen plus the J=83-82 line of $v$=0 HC$_7$N.}
\label{fg:hc5nvib}%
\end{figure}

\subsection{Vibrationally excited HC$_5$N}
\label{hc5n_vib}

The latest laboratory data on the $\nu_{11}$ and $\nu_{10}$ vibrational states of HC$_{5}$N, 
their overtones and their combinations, have been presented by Yamada et al. (2004). 
In our line survey, we have detected rotational lines in the following vibrationally 
excited states of HC$_{5}$N (most of them for the first time in space): $\nu_{11}$, 
2$\nu_{11}$, $\nu_{10}$, and 3$\nu_{11}$, at energies of 105, 210, 230, and 315 
cm$^{-1}$ respectively. Since 
these lines also should arise from the SEE analyzed in PAR04, we have just tried 
to find the HC$_{3}$N/HC$_{5}$N abundance ratio that best fits the observations, keeping 
the same SEE physical parameters of PAR04. The result is HC$_{3}$N/HC$_{5}$N $\sim$ 3$\pm$1 
(see Fig. \ref{fg:hc5nvib}). This result confirms previous estimates made from vibrational bands 
of both species seen in absorption by ISO around 15 $\mu$m (Cernicharo et al. 2001).  
Obviously, the fit cannot be perfect for so many lines. In 
particular, Fig. \ref{fg:hc5nvib} shows that some absorption exists between -50 and -40 
 km$\cdot$s$^{-1}$  for J$_{up}$=35 in the $\nu_{11}$ state that is not reproduced by the model. 
This can be due to a fraction of a higher velocity outflow that intersects the line of sight 
toward the continuum source that has not been considered in the model. This absorption 
component is otherwise clearly visible in the rotational lines inside the $\nu_{7}$ state 
of HC$_{3}$N, but does not affect the other lines shown in Fig. \ref{fg:hc5nvib}. 
The detections in the HC$_{5}$N vibrational states extend from J$_{up}$=31 to 
J$_{up}$=104 for $\nu_{11}$, J$_{up}$$\sim$87 for 2$\nu_{11}$, 
J$_{up}$$\sim$77 for $\nu_{10}$, and J$_{up}$$\sim$53 for 3$\nu_{11}$. The fact that for HC$_{3}$N we 
detect rotational lines from 
vibrational states a factor 3-4 higher in energy than for HC$_{5}$N is consistent, for 
the sensitivity of the survey, with the derived HC$_{3}$N/HC$_{5}$N 
abundance ratio, and the dipole moment and partition function ratios of both species.

\begin{figure}[t]
\includegraphics[angle=270,width=\columnwidth]{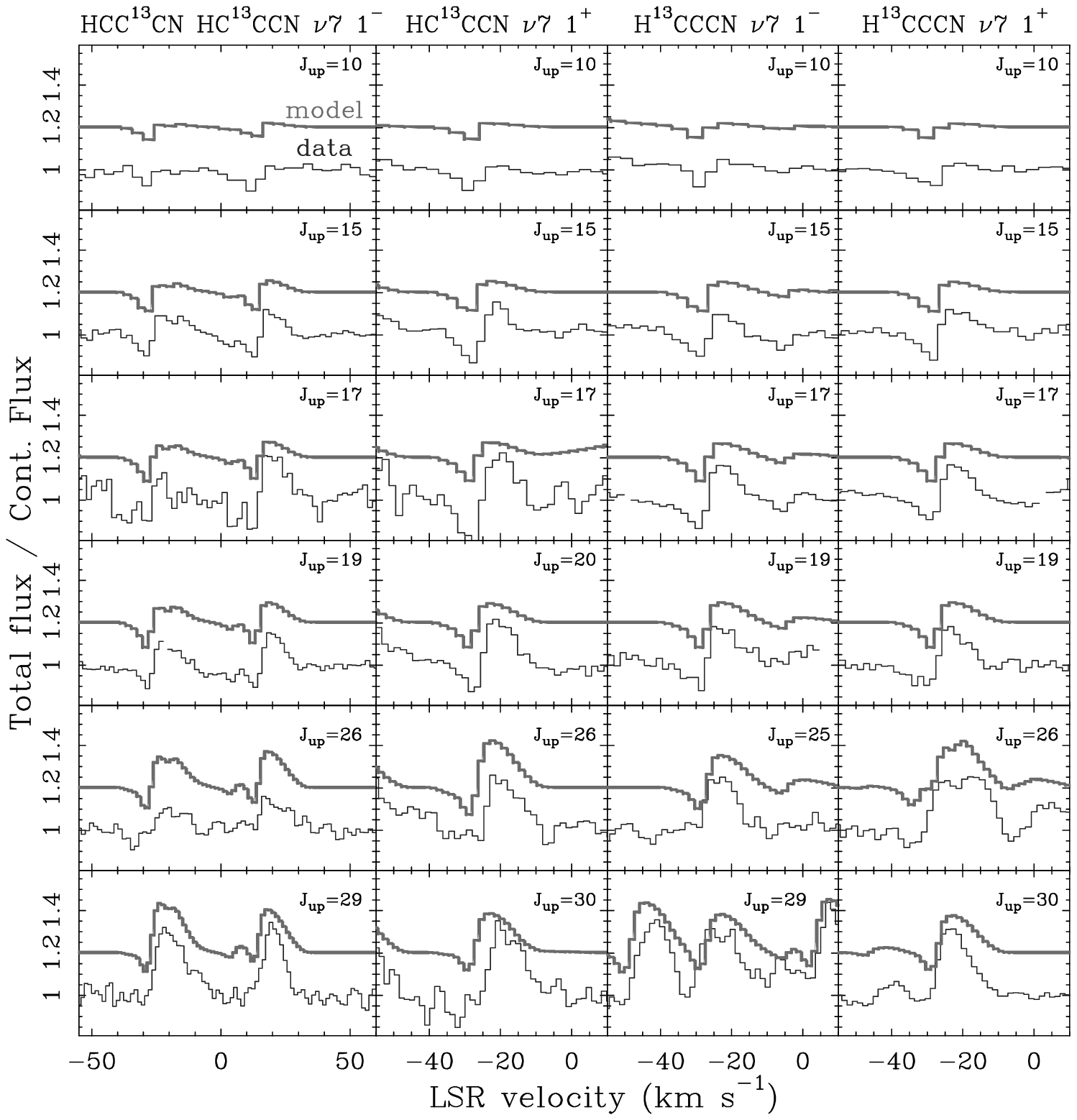}
\caption{Model predictions considering a $^{12}$C/$^{13}$C isotopic ratio of 15 in the slowly expanding 
envelope of CRL 618 {\bf (shifted by 0.2$\cdot$F$_c$)} compared to the observations for a sample of 
lines belonging to single $^{13}$C substituted 
HC$_{3}$N in the $\nu_{7}$ state. {\bf The velocity resolutions of model and data are the same
 ($\Delta v=c\cdot\frac{\Delta\nu}{\nu}$, with $\Delta\nu\simeq$1 MHz for all spectra 
and $\nu\simeq$J$_{up}\cdot$9.1 GHz}). Two weak features, between 
the HCC$^{13}$CN and HC$^{13}$CCN $\nu7$ 1$^-$ lines, clearly seen in the spectra with the best signal-to-noise ratio, 
correspond to the $\nu6$ 1$^+$ lines of both species. They have also been included in the model. All other extra features 
that appear in some spectra belong to vibrationally excited states of HC$_3$N.}
\label{fg:1213see}
\end{figure}

\subsection{$^{12}$C/$^{13}$C ratio in the SEE}
\label{1213ratio_see}
The $^{12}$C/$^{13}$C isotopic ratio in the SEE can be derived using HCC$^{13}$CN, 
HC$^{13}$CCN and H$^{13}$CCCN in the $\nu_{7}$ state. The $\nu_{6}$ state is also detected 
but the signal-to-noise ratio is too low for any estimation. Figure \ref{fg:1213see} shows 
the results of the SEE model for ($^{12}C$/$^{13}C$)=15 (PAR04). We estimate an error bar for 
this value of $\sim \pm$2. Kahane et al. (1992) gave lower limits for this ratio of 18 and 3.2 
depending on which transition (J=1-0 or J=2-1) of both $^{12}$CO and $^{13}$CO was used. Our 
estimate gives a much better constraint, and restricted to the SEE region, because it relies 
upon tens of lines without the saturation problems of CO. 


\subsection{Ground state HC$_5$N}
\label{hc5n_ground}
Using the result found in \ref{hc5n_vib}, we have run the model for the detected rotational 
lines of HC$_5$N in its ground vibrational state. 
From these calculations it becomes clear that 
the SEE alone cannot account for these lines. A significant contribution from gas at lower 
temperatures must exist. {\bf In the individual lines, there is no significant evidence of the}  
high velocity wind (HVW) seen in the 
most abundant molecules (see below). Therefore, we have expanded the model presented in PAR04 
to include a cold circumstellar shell (CCS), created during the AGB phase 
(see figure \ref{fg:crl618_model}). We have not simply considered a spherical geometry for this 
cold component in order to account for the optical/infrared images of the CSE that show  
that the envelope has a bipolar structure. This was also pointed out for 
the molecular gas in CER89 and Neri et al. (1992).   
 The parameters that define the CCS in our model are:

\begin{figure}[t]
\includegraphics[angle=0,width=\columnwidth]{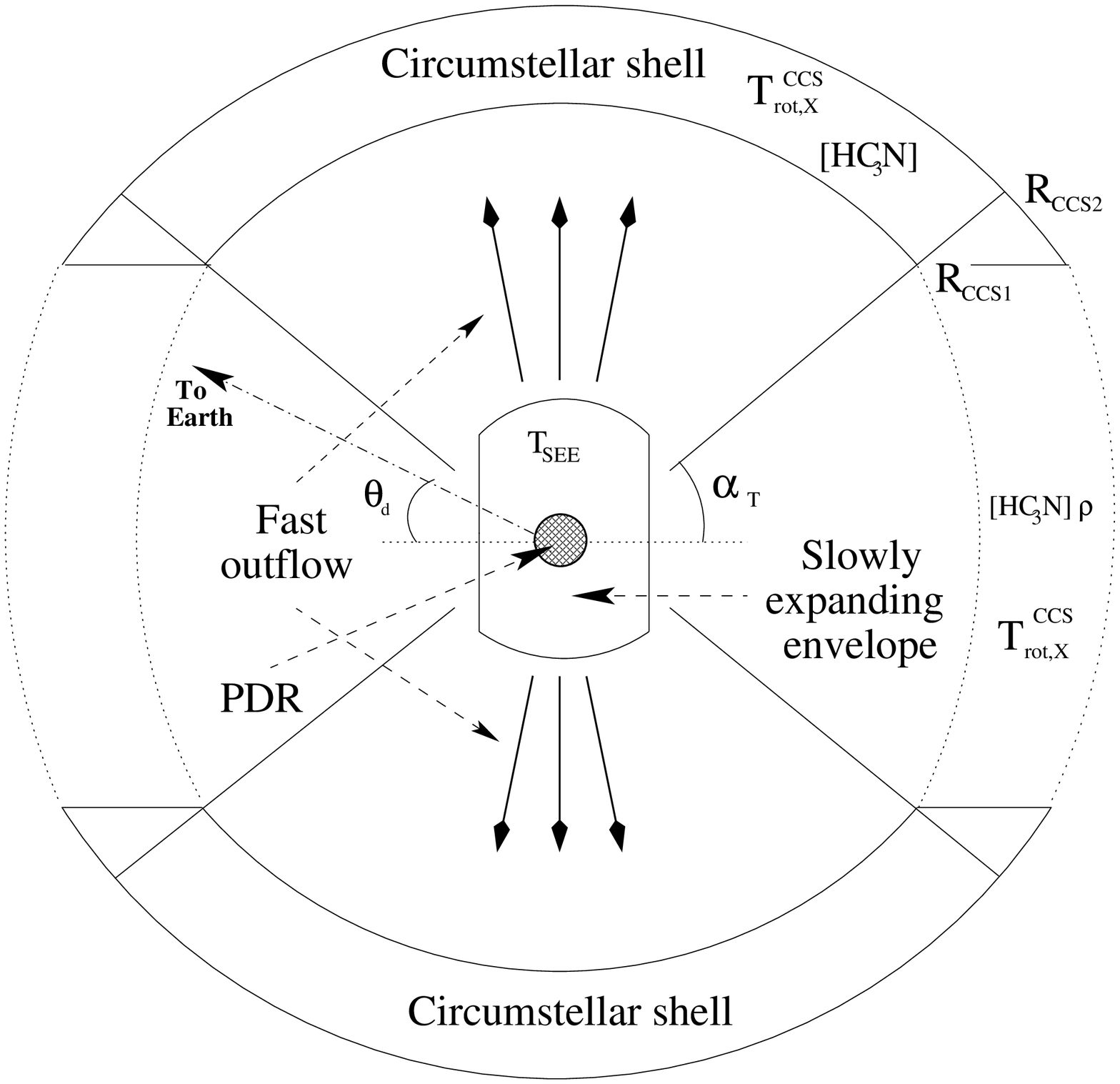}
\caption{Model of CRL 618 including a cold circumstellar shell first introduced in our work to 
reproduce the observed rotational lines of HC$_5$N in its ground vibrational state. {\bf Details in 
section \ref{hc5n_ground}.}}
\label{fg:crl618_model}
\end{figure}

\begin{itemize}
\item $R_{CCS1}$ and $R_{CCS2}$: Inner and outer boundaries of the shell (angles in arcseconds).

\item $\alpha_T$: Truncation angle (degrees), used to take into account the fact that this shell needs 
to be divided into at least two zones because the molecular density is  
probably larger in the polar regions along the axis of the high velocity wind. This fact is 
revealed by the ratio between emission and absorption in the profiles of the analyzed 
$v$=0 HC$_5$N rotational lines. 

\item T$^{CCS}_{rot,X}$: Rotational temperature (K) for species X in the CCS.

\item $[$HC$_3$N$]$ (at $R$=$R_{CCS1}$): 
Number density 
of HC$_3$N at the inner 
part of the shell for $\alpha > \alpha_T$ (for 
$\alpha < \alpha_T$ the density will by multiplied by a factor $\rho_{ccs}$). HC$_3$N will 
always be our reference molecule for the models 
of CRL 618. In each region we will consider a constant ratio with respect to it for 
each molecule. 

\item d$_{CCS}$: The radial distribution of HC$_{3}$N abundance is considered to vary as
[HC$_{3}$N$_{R}$] = [HC$_{3}$N$_{R_{CCS1}}$]($R$/$R_{CCS1}$)$^{d_{CCS}}$
in the CCS.

\item v$_{CCS}$ and v$_{turb,CCS}$: radial and turbulence velocity of the gas in the CCS.

\end{itemize}

\begin{figure*}[t]
\includegraphics[angle=270,width=\textwidth]{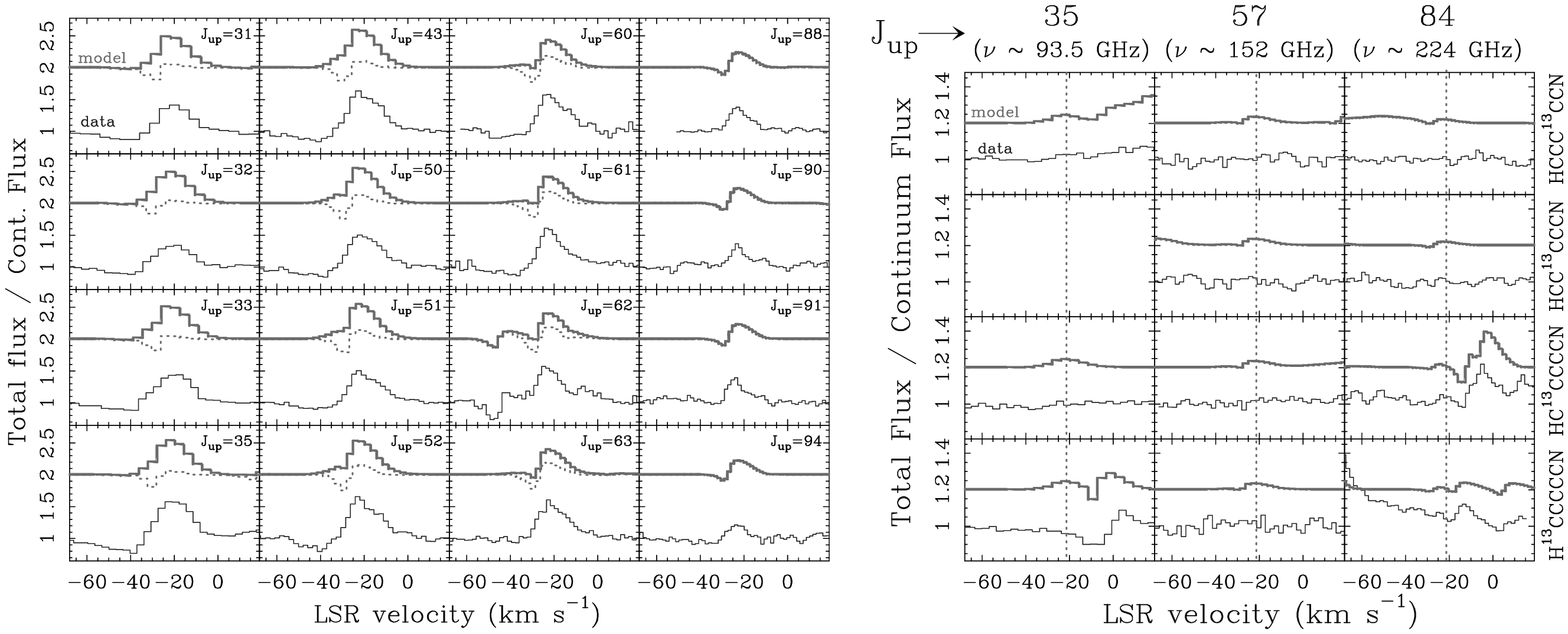}
\caption{(Left) Observed rotational lines of HC$_5$N in the ground vibrational state compared with the SEE model 
assuming a HC$_3$N/HC$_5$N value of 3 {\bf (dotted histogram, shifted by 1.0$\cdot$F$_c$) and with the expanded 
model (solid histogram, shifted by 1.0$\cdot$F$_c$), that includes the CCS, presented in this paper 
(Sect. \ref{results},  Fig. \ref{fg:crl618_model} and Table \ref{tb_CSpar})}. 
{\bf Both models are basically coincident for J$_{up}$ above $\sim$ 80}. 
(Right) Results of the CRL 618 model including the SEE 
and the CCS regions ({\bf solid histogram shifted by 0.2$\cdot$F$_c$}) 
for a sample of lines of $^{13}$C substituted HC$_5$N, compared to the observations. The 
simulations consider a $^{12}$C/$^{13}$C value of 15 in both regions. {\bf Velocity resolution of data and model 
are the same for each plot, as explained in previous figures.}}
\label{fg:hc5n_v0}
\end{figure*}

Just as we did in PAR04 for the physical parameters of the SEE, we had to run a considerable 
number of models to find a good fit of the ground state HC$_5$N lines. Again, the key 
parameters are the number density, the temperature  and the position of the CCS. The 
position has been fixed between {\bf diameters} 3.0$''$ and 4.5$''$  
after trying several other positions in our models.  Then, 
the temperature and density profiles can be  
established by running a grid of models. All the other parameters are then explored 
to fine tune the shape of the line as J$_{up}$ varies. For example, one of the most remarkable 
differences in the line profiles from ground to vibrationally excited HC$_5$N 
is the position of the maximum depth of the absorption, its width and the 
absorption/emission ratio. As shown in Figure \ref{fg:hc5nvib} and table 
\ref{tb:lineparameters}, the maximum depth of the 
absorption for vibrationally excited HC$_5$N happens at $\sim$ -30 km$\cdot$s$^{-1}$ in the 
2 and 3 mm windows {(as for the lines in vibrationally excited states of HC$_3$N presented 
in PAR04)}, whereas for the lines in the vibrational ground state it happens at 
around -40 km$\cdot$s$^{-1}$ and is generally wider and less deep.  
The observational differences have allowed us to 
determine the kinematic conditions in the CCS. The parameters of 
the best simulation (Fig. \ref{fg:hc5n_v0}) are shown on Table \ref{tb_CSpar}. The rotational 
temperature is in the range 50-70 K. In fact, a simulation with T$_{rot}$=50 K provides a better 
fit for lines below J$_{up}\sim$55, whereas T$_{rot}$=70 K provides a better fit for the range 
J$_{up}\sim$55-80. The simulation shown in Fig. \ref{fg:hc5n_v0} is a compromise 
between the two cases (T$_{rot}$=60 K). Above J$_{up}\sim$80, the line profiles are 
almost exclusively determined by the SEE region. The error bar for the 
$[$HC$_3$N$]$/$[$HC$_5$N$]$ = 3 ratio in the CCS 
can again be estimated to be $\pm$1.  
The same absorption at v$_{LSR}<$-40  km$\cdot$s$^{-1}$  reported in section \ref{hc5n_vib}, and 
not perfectly modeled here, is seen in some $v$=0 HC$_5$N spectra. Absorptions 
centered at $\sim$ -40 km$\cdot$s$^{-1}$ have previously been reported in this object in molecules 
such as CO (Herpin et al. 2002), HCN (CER89 and Neri et al. 1992), HCO$^+$ 
(CER89 and S\'anchez-Contreras \& Sahai, 2004), HC$_3$N (CER89, PAR04), and CS 
(Hajian et al. 1995). There is general agreement to 
assign that absorption to the AGB circumstellar envelope.

S\'anchez-Contreras \& Sahai (2004) suggest that 
an elongated structure of $\sim$11'' diameter  
with expansion velocity of $\sim$17.5 km$\cdot$s$^{-1}$ is responsible 
for part of the emission of the low velocity molecular gas in 
their HCN J=1-0 and HCO$^+$ J=1-0 interferometric maps. This 
structure is most probably {\bf an extension (due a larger abundance of this molecules 
in cold regions) of}  the CCS we are modeling here for single 
antenna data. The difference between this larger shell and our 
CCS is probably due to the different excitation conditions of the low-J 
lines of HCN and HCO$^+$ and the high-J lines of HC$_5$N explored in this work.

 On the basis of the extended model, that now includes two gas regions, we can check 
for the presence of longer cyanopolyynes rather easily because only their abundance ratio 
with respect to HC$_5$N or HC$_3$N is needed to make predictions of their lines and then check 
against the data (assuming the same rotational temperature). This is presented in the 
following subsections.

\begin{table}
\caption[]{Model parameters of the Circumstellar Shell (see figure \ref{fg:crl618_model}) that 
provide the best fit to the HC$_5$N rotational lines in the ground vibrational state (see 
figure \ref{fg:hc5n_v0}).}
\begin{center}
\begin{tabular}{lrc}
parameter & value  \\
\hline
2$\cdot R_{CCS1}$ (arcsec) & 3.0 \\
2$\cdot R_{CCS2}$ (arcsec) & 4.5 \\
$\theta_{d}$ (deg) &  40 \\
$\alpha_T$ (deg) &  30 \\
T$^{CCS}_{rot,HC_{5,7}N}$ (K) & 50-70 \\
$[$HC$_3$N$]$ (at $R$=$R_{CCS1}$) (cm$^{-3}$) & 0.8 \\
d$_{CCS}$ (no units) & -1.0 \\
$\rho_{CCS}$ (no units) &  1/4   \\
$[$HC$_3$N$]$/$[$HC$_5$N$]$ & 3 \\
$[$HC$_5$N$]$/$[$HC$_7$N$]$ & 6 \\
v$_{CCS}$  (at $R$=$R_{CCS1}$) (km$\cdot$s$^{-1}$) & 22  \\
v$_{turb,CCS}$ (at $R$=$R_{CCS1}$) (km$\cdot$s$^{-1}$) & 12  \\
v$_{CCS}$  (at $R$=$R_{CCS2}$) (km$\cdot$s$^{-1}$) & 8 \\
v$_{turb,CCS}$ (at $R$=$R_{CCS2}$) (km$\cdot$s$^{-1}$)& 5 \\
\end{tabular}
\end{center}
\label{tb_CSpar}
\end{table}

\subsection{Isotopic HC$_5$N and $^{12}$C/$^{13}$C ratio in the CCS} We have looked at the possible 
detection of the $^{13}$C substituted species of HC$_5$N by comparing the observations 
at the frequencies of their rotational lines in the ground vibrational state with the extended  
model results considering a $^{12}$C/$^{13}$C isotopic ratio of 15 in the SEE (as derived in 
section \ref{1213ratio_see}) and also in the CCS. The simulations indicate that this value 
is an upper limit for the CCS. This is important to later evaluate the HCN and HNC abundances from 
H$^{13}$CN and $^{13}$C (section \ref{sct:hcn_hnc}).

\subsection{HC$_3$N in the CCS}
\label{hc3nccs}
\begin{figure}[t]
\includegraphics[angle=270,width=\columnwidth]{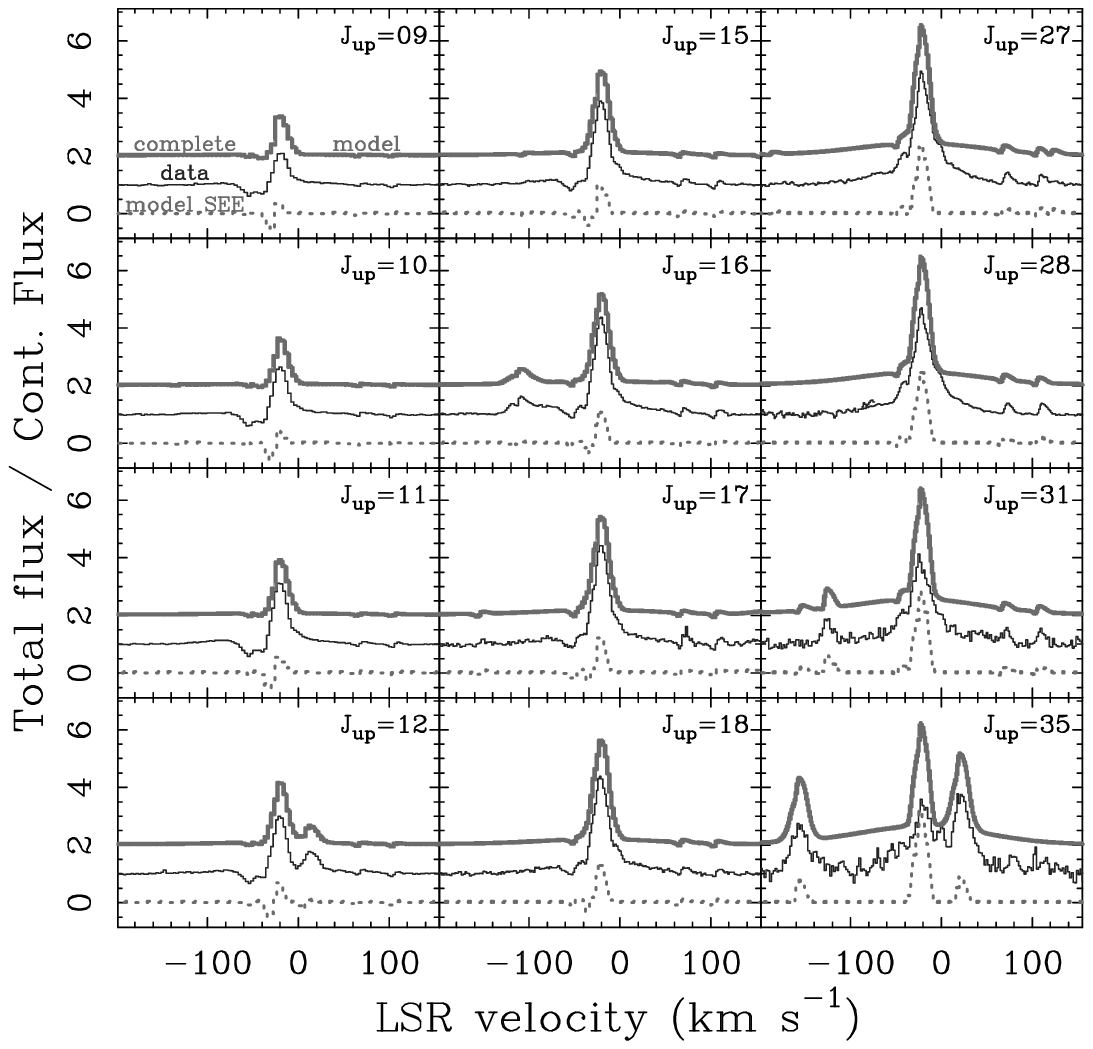}
\caption{Observed rotational lines of HC$_3$N in the ground {\bf vibrational} state compared with 
the SEE+CCS model assuming a HC$_3$N/HC$_5$N value of 3, plus a HVW component according to section \ref{hc3nccs}
{\bf (solid histogram shifted by 1.0$\cdot$F$_c$). The model for the SEE region alone is also plotted for comparison 
(dotted line, shifted by -1.0$\cdot$F$_c$)}. Some 
other lines appear in the data and are also modeled. Most of them are rotational transitions within different 
vibrational states of HC$_{3}$N, HC$_5$N or their isotopologues, but there are some belonging to other species 
the discussion of which we leave for a separate work: H$_2$CO and C$_5$H in J$_{up}$=16, 
and c-C$_3$H$_2$ in J$_{up}$=35.  
All observations have been carried out with the IRAM-30m telescope except J$_{up}$=31,35 (CSO). {\bf Velocity 
resolutions of data and model are the same as explained in previous figures.}}
\label{fg:hc3n_v0}
\end{figure}

The HC$_3$N lines in vibrational states below 2$\nu_7$ show an emission that exceeds what is 
predicted by the SEE model. Including the CCS component as in section \ref{hc5n_ground} provides 
a first approximation to the detected signal (see figure \ref{fg:hc3n_v0}) that is in fact quite 
good for J$_{up}$ below $\sim$20. However, the HVW component is also visible in the high-J $v$=0  
lines and therefore should be included in the model. The high velocity outflow is modeled by just 
considering a gas component with a given excitation temperature (350 K), average column density 
(2.5$\cdot$10$^{17}$ cm$^{-2}$), filling factor at a reference frequency (1.3$\cdot$10$^{-3}$ 
at 90 GHz) and halfwidth velocity outflow (150 km$\cdot$s$^{-1}$). The rather high temperature for this 
component is confirmed by the dominant role that the broad HVW component has at high J's (see CSO 
lines in figure \ref{fg:hc3n_v0}). More precise details 
about the structure of the fast outflow are difficult to obtain with the present data. A better constraint  
on the temperature should be obtained by observing a few $v$=0 HC$_3$N lines above 400 GHz (on going).

\subsection{HCN and HNC emission} 
\label{sct:hcn_hnc}
At this point of the analysis we have a model that considers the three main regions of the gas emission 
in the millimeter wave spectrum of CRL618: SEE, CCS, and HVW. We can try to fit the observed lines 
from other species by adjusting only 
their ratio to HC$_3$N in the assumption that the rotational temperature is the same. This approximation 
will be quite good for those species arising mainly from the SEE, due to the high density of the region 
(populations are thermalyzed). It could also be useful for the HVW. However, the low densities imply 
that the full analysis for all molecular species will be more complex in the CCS.

\begin{figure}[t]
\includegraphics[angle=0,width=10cm]{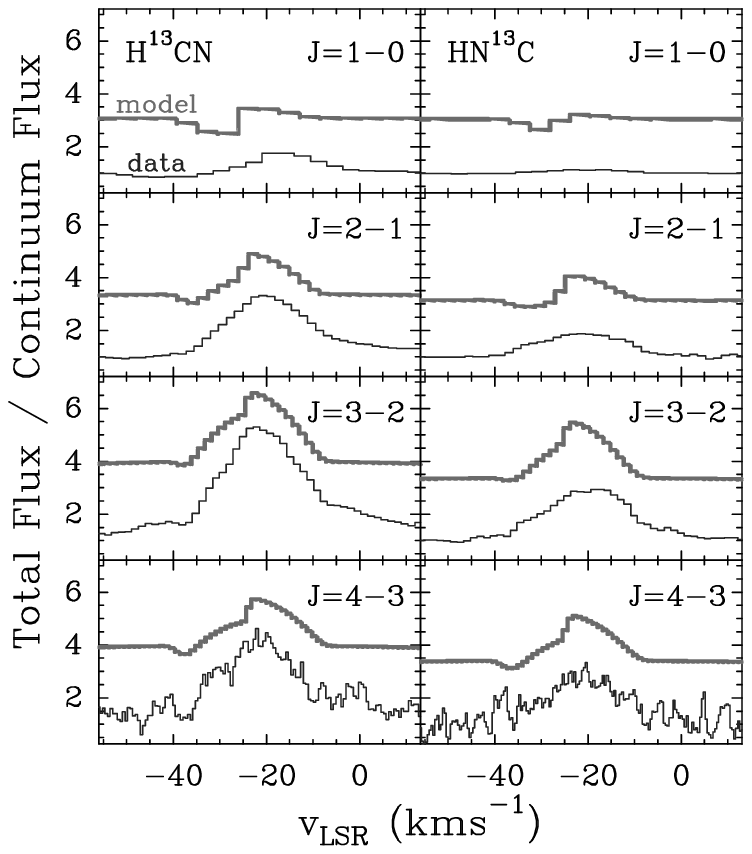}
\caption{Observed rotational lines of H$^{13}$CN and HN$^{13}$C compared with the SEE+HVW model 
{\bf (shifted by 2.0$\cdot$F$_c$)} 
assuming  H$^{13}$CN/HC$_3$N and HN$^{13}$C/HC$_3$N values of 1/2 and 2/15 respectively. All observations 
have been carried out with the IRAM-30m telescope except J$_{up}$=4 (CSO). The very low excitation temperatures of 
J$_{up}>$=1,2 levels, compared to those in the SEE and HVW regions, results in the the poor agreement between data and  
model for the lines arising from those levels. Nevertheless, the other lines provide a useful {\bf upper limit 
(the CCS contribution should be important)} about the HCN and HNC abundances in the SEE and HVW regions. 
{\bf Velocity resolutions of model and data are the same for the different lines, as explained in previous figures.}}
\label{fg:h13cn_hn13c}
\end{figure}

In this work we will focus only on HCN and HNC. For opacity reasons, we cannot probe the inner SEE and HVW regions 
using the main isotopologues of both species, specially as J increases. Instead we can use H$^{13}$CN and HN$^{13}$C 
assuming a SEE  
$^{12}$C/$^{13}$C value of 15 (see above) {\bf but even for these isotopologues some contribution from the CCS can 
be expected. As a result, our simulations considering only the SEE+HVW (see figure \ref{fg:h13cn_hn13c}) 
for J$_{up}>$2 provide upper limits for the H$^{13}$CN/HC$_3$N and HN$^{13}$C/HC$_3$N ratios in the SEE. The values 
obtained are  about 1/2 and 2/15 respectively which translate into HCN/HC$_3$N $\sim$ 7.5 and HCN/HC$_3$N $\sim$ 2.0}. 

\begin{figure}[t]
\includegraphics[angle=0,width=\columnwidth]{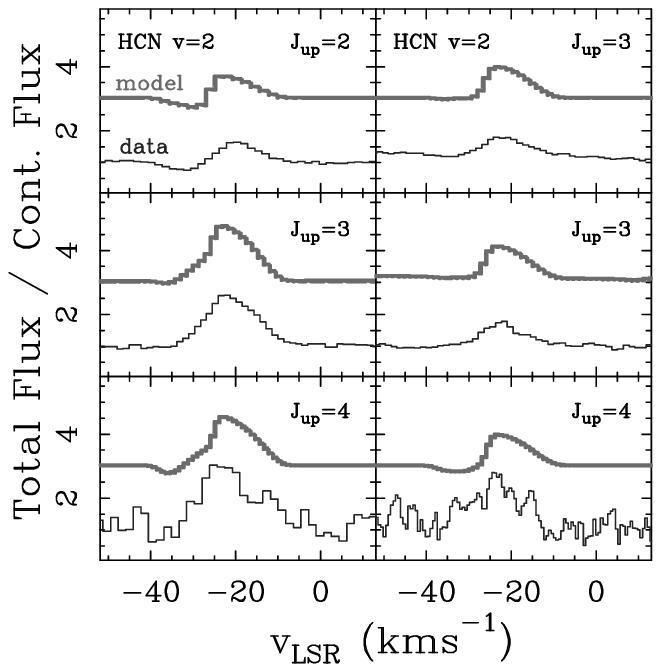}
\caption{Observed rotational lines of HCN and HNC {\bf in the $v$=2 vibrational state}  
compared with the SEE model {\bf (shifted by 2.0$\cdot$F$_c$)} assuming  HCN/HC$_3$N and HNC/HC$_3$N values 
of 2 and {\bf 1/5} respectively. All observations have been carried out with the IRAM-30m telescope 
except J$_{up}$=4 (CSO). {\bf Velocity resolutions of data and model are the same as in previous figures.} }
\label{fg:vibhcnhnc}
\end{figure}

{\bf To improve the above estimates in the SEE it is better to use vibrationally excited states of HCN and HNC since the 
CCS contribution should be negligible for them}. The lowest one ($v_2$=1), at 712 and 477 cm$^{-1}$ respectively for HCN and HNC, is 
detected in both molecules ($l$ 
doublets for all rotational transitions except the forbidden J=1-0). We have selected those lines that have been 
observed and are not badly blended with other species to get a better constraint for the above ratios. This analysis 
(Figure \ref{fg:vibhcnhnc}) results on HCN/HC$_3$N $\sim$ 2 {\bf and HNC/HC$_3$N $\sim$ 1/3 to 1/5}. It is thus 
demonstrated that much lower ratios of HCN and HNC with respect to longer cyanopolyynes are found in the inner gas of CRL618 
(PPNe phase) compared to those found in IRC+10216 (AGB phase). See discussion in section \ref{chemicalmodels}.


\subsection{Ground state HC$_7$N}

\begin{figure}[t]
\includegraphics[angle=0,width=10cm]{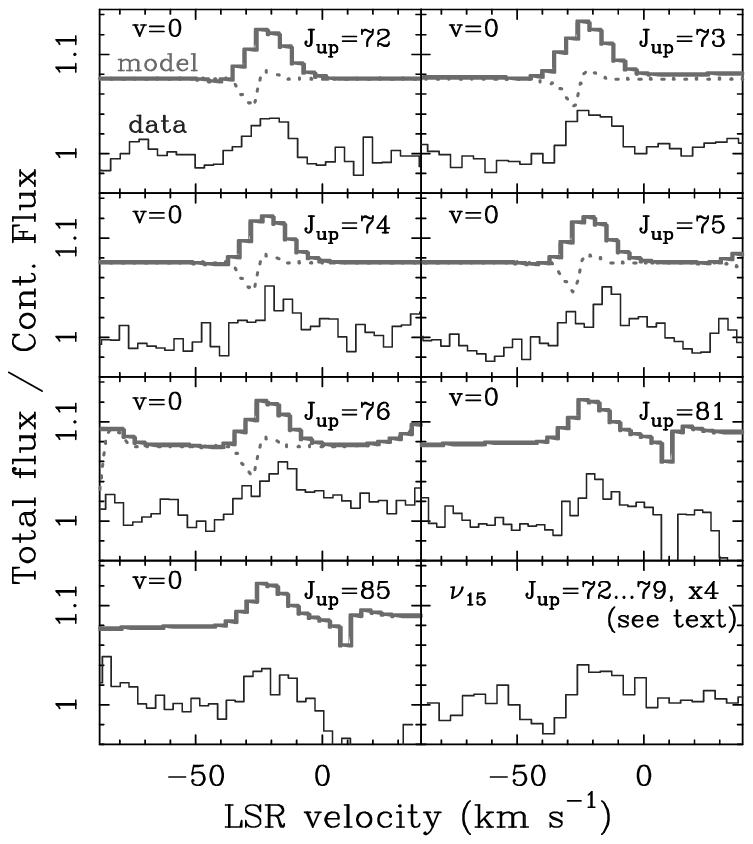}
\caption{HC$_7$N ($v$=0) lines observed toward CRL618 compared with model results including 
the SEE and CCS regions (solid {\bf histogram shifted by 0.075$\cdot$F$_c$}) and the CCS region alone 
(dotted line {\bf also shifted by 0.075$\cdot$F$_c$}). {\bf The considered value of  
HC$_{5}$N/HC$_{7}$N} is 3, which provides a very good fit. The spectrum in 
the lower-right corner box is the addition of the $\nu_{15}$ HC$_7$N rotational lines listed in 
table \ref{tb:hc7nvib} compared to the model results for both the CCS and the SEE. A 4.0 
scaling factor has been introduced to keep the same scale of all other spectra. } 
\label{hc7n_vib0}
\end{figure}

Finally, we have studied the HC$_{7}$N rotational lines within the $v$=0 state. This molecule 
is clearly detected since all 
transitions between J$_{up}$=72 and J$_{up}$=85 appear above a 3$\sigma$ level  
except for three cases due to a coincidence with a much stronger line (see Table \ref{tb:hc7nvib} 
and figure \ref{hc7n_vib0}). Note that most of the HC$_7$N $v$=0 signal arises from the CCS. We reach this  
conclusion after running models for which no HC$_{7}$N was present in the CCS, and its ratio to HC$_{5}$N 
was set to be 1/3 in the SEE. In this case, the model predicts some detectable absorption 
and almost no noticeable emission.   
The fact that HC$_{7}$N is seen almost exclusively in emission 
with a much larger intensity than in this model, can be easily reproduced if the CCS is included 
with a HC$_{5}$N/HC$_{7}$N abundance ratio around 3 and the same rotational temperature for both 
species  (see figure \ref{hc7n_vib0}). The model also 
predicts that above J$_{up}$=85 (at an energy level of 134 cm$^{-1}$) the signal would be 
too week to be detected above the 3$\sigma$ level of the survey. Similarly, the intensity 
of lines of HC$_{9}$N, with a ratio of 1/3 or less with respect to HC$_{7}$N, is estimated to be well 
below the detection limit of our observations.

\begin{table}
\caption[]{Detected lines of HC$_7$N toward CRL 618 (laboratory frequencies in GHz are rounded 
to 0.1 MHz). {\bf Frequencies} in parenthesis correspond to {\bf co-added lines} in the velocity 
scale for a tentative detection of $\nu_{15}$ (see text). }
\label{tb:hc7nvib}
\begin{center}
\begin{tabular}{crrr}
   & \multicolumn{3}{c}{Vibrational state of HC$_7$N} \\ 
\cline{2-4}
 \multicolumn{1}{c}{J$_{up}$}  &    $v$=0    & {$\nu_{15}$ 1$^-$} & {$\nu_{15}$ 1$^+$}  \\ 
 \multicolumn{1}{c}{72}    &    81.2101       &    (81.3115)       &    (81.3630)       \\
 \multicolumn{1}{c}{73}    &    82.3379       &          &    (82.4929)       \\
 \multicolumn{1}{c}{74}    &    83.4656      &           &           \\
 \multicolumn{1}{c}{75}    &    84.5933        &  (84.6989)         &           \\
 \multicolumn{1}{c}{76}    &    85.7211       &  (85.8281)         &   (85.8824)        \\
 \multicolumn{1}{c}{77}    &           &   (86.9572)        &   (87.0122)        \\
 \multicolumn{1}{c}{78}    &    87.9765     &       &  (88.1421)        \\
 \multicolumn{1}{c}{79}    &    89.1042         &   (89.2154)        &   (89.2719)        \\
 \multicolumn{1}{c}{80}    &           &          &          \\
 \multicolumn{1}{c}{81}    &    91.3596       &           &           \\
 \multicolumn{1}{c}{82}    &    92.4873       &           &           \\
 \multicolumn{1}{c}{83}    &                  &           &           \\ 
 \multicolumn{1}{c}{84}    &    94.7426       &           &           \\ 
 \multicolumn{1}{c}{85}    &    95.8703       &           &           \\ 
 \multicolumn{1}{c}{86}    &                  &           &           \\ 
 \multicolumn{1}{c}{87}    &                  &           &           \\
 \multicolumn{1}{c}{88}    &                 &           &           \\ 
\end{tabular}
\end{center}
\end{table}

\subsection{Vibrationally excited HC$_7$N}
The frequencies of the rotational transitions in the lowest energy bending modes of 
HC$_7$N ($\nu_{13}$,$\nu_{14}$,$\nu_{15}$) have become available in a recent 
work  by Bizzochi \& Degli Esposti (2004). We have searched for those in the lowest 
vibrational state (0,0,1) but no individual detection has been obtained. These 
lines have been modeled by considering computed ab initio values for the (0,0,1), (0,1,0) 
and (1,0,0) energies: 62, 163 and 280 cm$^{-1}$ respectively (Botschwina et al. 1997). A 
tentative detection is obtained when performing the addition in the v$_{LSR}$ scale  
of the observed signal at the positions of the 1$^-$ and 1$^+$ J$_{up}$=72 to 79 lines 
(see figure \ref{hc7n_vib0}). Five lines in the considered J$_{up}$ range were discarded 
from the summation because of blending with stronger lines (see Table \ref{tb:hc7nvib}).

\section{Comparison with chemical models}
\label{chemicalmodels}

The abundance ratio for consecutive members of the cyanopolyynes family of molecules 
in IRC+10216 is $\sim$ 3 (Cernicharo et al. 1987). However, the 
HCN/HC$_3$N ratio in the same object is larger than 100. In CRL 618 our results indicate 
also an abundance ratio near 3 for HC$_3$N/HC$_5$N and $\sim$6 for HC$_5$N/HC$_7$N, but for HCN/HC$_3$N 
it is around 2-3. The difference in this ratio at different evolutionary stages of C-rich objects 
tells us about the chemical processing of the gas when the central star evolves toward 
the white dwarf stage. CER04 has modeled the role of the strong UV field arising from 
the hot central star on the chemical abundances of the polyacetilenes, cyanopolyynes and 
carbon clusters. This work shows that HCN is quickly photodissociated allowing the production 
of CN which reacts with C$_2$H$_2$ and C$_2$H to produce HC$_3$N and longer cyanopolyynes, 
thus explaining the low HCN/HC$_3$N ratio found in CRL 618. Moreover, this behavior 
is similar to that found by Cernicharo et al. (2001) for C$_4$H$_2$ and C$_6$H$_2$. The 
large abundance of HNC with respect to HCN in CRL618 (HCN/HNC $\sim$ 10 instead of $>$ 200 in 
IRC+10216, Cernicharo et al. 1987) also points in the same direction. It 
seems that in a short period of time it could be possible to transform an important fraction 
of HCN into HC$_3$N and longer cyanide carbon chains.  

\section{Summary and future work}
\label{summary}

We have extended our previous analysis of the gas shells surrounding 
the protoplanetary nebula CRL618, by studying the pure rotational lines of 
HC$_{5}$N in its fundamental and the lowest 4 vibrationally excited states 
(first astronomical source in which vibrationally 
excited HC$_{5}$N has been detected), and HC$_{7}$N rotational lines in its 
fundamental vibrational state. We have found that the HC$_{3}$N/HC$_{5}$N 
ratio in the innermost slowly expanding envelope is $\sim$ 3. With this ratio,  
and all the physical parameters previously derived in PAR04, a good match to 
the observed HC$_{5}$N lines in vibrationally excited states is found. However, 
the predicted lines for the ground vibrational state are far too weak, suggesting 
that a more extended and colder gas shell has to be considered. The 
physical parameters of this envelope, called cold circumstellar shell (CCS) in this work, 
have been determined. The position of the shell is $\sim$3.0''-4.5'' from the central 
star and its excitation temperature is only $\sim$60 K (compared to 250-275 K in the 
innermost SEE). The column density of the CCS in front of the continuum source is 
negligible compared to that of the SEE. With the CCS derived parameters it is found 
that the observed HC$_{7}$N lines in the ground vibrational state can be well reproduced 
with a HC$_{5}$N/HC$_{7}$N abundance ratio of $\sim$ 3. The isotopic ratio $^{12}$C/$^{13}$C is found 
to be $\sim$ 15 in the SEE from the lines of vibrationally excited HC$_{3}$N. The same value 
must be an upper limit in the CCS according to the non detection of $^{13}$C substituted 
isotopologues of HC$_{5}$N.

The analysis of the $v$=0 HC$_{3}$N lines requires to introduce a third component, the high 
velocity outflow, although its structure cannot be precisely known. Finally, the ratios  
HCN/HC$_{3}$N and HNC/HC$_{3}$N have been derived using the first four rotational transitions of 
H$^{13}$CN and HN$^{13}$C, and some lines of vibrationally excited HCN. The result of a 
much lower HCN/HC$_{3}$N value than in IRC+10216 indicates a processing of HCN into longer cyanopolyynes 
in the protoplanetary nebula stage.

A first examination of lines from all other molecules detected in the survey shows that 
most of them can be well fitted using the model of the CCS, HVW and SEE gas regions described in 
this paper and in PAR04, so that a good estimate of their abundances in these regions can be 
achieved. Therefore, the next step in our analysis of the CRL618 line survey 
will be to provide a picture of the chemical content in its three main gas components.

\acknowledgments
We thank the support of the IRAM-30m staff during the long completion of the 
line survey. CSO operations are supported by U.S. NSF grant AST 22-09008. 
This work has also been supported by Spanish DGES and PNIE 
grants ESP2002-01627, AYA2002-10113-E and AYA2003-02785-E.

\end{document}